# Real-time Autonomous Control of a Continuous Macroscopic Process as Demonstrated by Plastic Forming

Shun Muroga[1]*, Takashi Honda[2], Yasuaki Miki[1], Hideaki Nakajima[1], Don N. Futaba[1], Kenji Hata[1]

1: Nano Carbon Device Research Center, National Institute of Advanced Industrial Science and Technology, Tsukuba Central 5, 1-1-1, Higashi, Tsukuba, Ibaraki, 305-8565, Japan

2: Research Association of High-Throughput Design and Development for Advanced Functional Materials (ADMAT), Tsukuba, Ibaraki, 305-8568, Japan

*Corresponding author. E-mail: muroga-sh@aist.go.jp

**Abstract**

To meet the demands for more adaptable and expedient approaches to augment both research and manufacturing, we report an autonomous system using real-time *in-situ* characterization and an autonomous, decision-making processer based on an active learning algorithm. This system was applied to a plastic film forming system to highlight its efficiency and accuracy in determining the process conditions for specified target film dimensions, importantly, without any human intervention. Application of this system towards nine distinct film dimensions demonstrated the system ability to quickly determine the appropriate and stable process conditions (average 11 characterization-adjustment iterations, 19 minutes) and the ability to avoid traps, such as repetitive over-correction. Furthermore, comparison of the achieved film dimensions to the target values showed a high accuracy ($R^2$ = 0.87, 0.90) for film width and thickness, respectively. In addition, the use of an active learning algorithm afforded our system to proceed optimization with zero initial training data, which was unavailable due to the complex relationships between the control factors (material supply rate, applied force, material viscosity) within the plastic forming process. As our system is intrinsically general and can be applied to any most material processes, these results have significant implications in accelerating both research and industrial processes.

## 1. Introduction

All things in daily life possess a form or shape, which is determined based on its function, such as for ease of handling, motion, or construction. Such macroscopic forms and their manufacturing processes are intrinsically connected to diverse industry sectors including plastics, rubber, food, and pharmaceuticals. In various systems, slurries, pastes, and high-temperature melts are precisely controlled and engineered in macroscopic forming processes to produce diverse commonplace products in mass-quantities, for example tapes, rods, and sheets (**Figure 1**a). Thus, these macroscopic forming processes play a vital role in the creation of everyday products. However, with the increasingly changing needs of the user, the target properties consequently change. This demands a more adaptable approach to fabrication and process control. Thus, the manual adjustment and "tuning" of the manufacturing processes which entail a wide range of parameters can be labor and time intensive. Against this backdrop, significant effort has been placed in research to advance from automated process control to *autonomous* process control.

Autonomy is a concept one step beyond the conventional idea of automation, as it incorporates the ability for independent decisions-making, which, in principle, completely removes the need for human intervention. This paradigm frees the human "operators" to focus attention on other scientific or industrial issues, particularly those not suited to automation or autonomous systems. Recently, research into the development of autonomous systems, i.e., "*autonomization,*" has been intensively pursued worldwide especially in the field of materials exploration, such as organics, inorganics, and nanomaterials. In the case of organics, there have been reports on optimizing chemical synthesis in batch[1] and flow reactors[2–4], as well as the exploration of organic solar cells[5], organic semiconductors[6], photocatalysts[7], and photo-curable resins[8]. For inorganics, the application of autonomous systems has been reported involving thin-film synthesis[9,10] and combustion synthesis[11], along with the investigation of magnetic materials[12] and phase-change memory materials[13]. Regarding nanomaterials, autonomization has been applied to the chemical vapor deposition synthesis of carbon nanotubes for investigating such aspects as growth rates and physical structures [14,15] Furthermore, the use of autonomization has also been applied to the exploration of process protocols. For example, Attia *et al*. successfully performed autonomous exploration to devise a six-step charging protocol that improved battery longevity.[16] In addition, Nagai *et al*. autonomously identified a stepwise heating protocol that reduced defects during the drying process of fuel cell slurries [17]. These examples demonstrate the diversity in the applicability of autonomous control in the various research fields to accelerate research.

In contrast, autonomization for macroscopic forming processes has not received significant attention, and the research in this area is surprisingly underdeveloped given its significance to manufacturing. This may be, in part, due to the difficulty in integrating several separate disciplines, such as precision automation, autonomous control, and system integration. While some research has focused on autonomous batch forming processes using 3D printers [18,19], no research has been conducted on the autonomization of continuous macroscopic forming, despite its direct applicability to numerous continuous synthesis processes as well as its paramount importance for industrial mass production. To achieve autonomous control for a continuous forming process, a "closed-loop" system equipped with an autonomous decision-making process and control based on real-time characterization data, such as dimensional and/or property characterization is required as the process does not halt.

Here, we report an autonomous system for a continuous macroscopic process using real-time *in-situ* characterization feedback (**Figure 1**b). As a demonstration, we applied our autonomous system to the plastic forming into films to highlight its efficiency and accuracy in achieving diverse target dimensions without human intervention. In this study, our system controlled the entire plastic forming process, where plastic pellets are fed, melted, transported, and extruded through a die (i.e., slit), cooled and collected onto a roll[20,21]. The system was also equipped with real-time *in-situ* thickness and width monitoring to appropriately balance the various process parameters (material feed rate, draw rate, and applied heat), as illustrated in **Figure 1**c. Because of the complex relationships between these factors, the balancing of these factors was critical for not only determining conditions for targeted film dimensions, but also to ensure film uniformity[22]. In addition, our system showed "instability avoidance" (i.e., ability to avoid being trapped within a parameter space of unstable states). Accuracy and efficiency of this system were demonstrated by the rapid convergence for nine diverse target values spanning different combinations of widths and thicknesses. Importantly, the decision-making section of our system was based on an active learning algorithm, which required no prior training, and target values could be achieved in an average of 11 iterations (~19 minutes). As this system is intrinsically general, it can be applied to any type of material process, and thus we believe that our findings will significantly promote the autonomization of continuous forming processes, paving the way for broader applications and implementations in the numerous research and industrial fields.

## 2. Results and Discussion
### 2.1 Demonstration of Autonomous Forming Process

To begin, we demonstrate the ability of our system to efficiently and independently determine the conditions to fabricate a film of specified dimensions (width/thickness) by employing an active learning processer to guide the autonomous operation. The sequence of states (i.e. "trajectory") following the progress is shown in **Figure 1**d. The progression of process adjustments is depicted as a transition from blue to red, with red indicating the endpoint. Star-shaped markers indicate the target positions for width and thickness, and the circular markers represent the states during the progression of the convergence process. In viewing this trajectory, we can make three observations. First, the conditions to fabricate the film with the target width and thickness were achieved in eight characterization-adjustment iterations, which required only 24 minutes to complete. This highlights one of the features of this system, process efficiency, which would be challenging to achieve through manual operation. This achievement is particularly exceptional because the system was not provided with any prior training and without using preset responses as used in commercial air-conditioning systems. Second, the trajectory appears incomprehensible to human understanding as it appears to be completely random in nature. However, by calculating and plotting relative root mean square at each iteration, the clear reduction in error indicates the apparent gradual target convergence, which is interrupted twice with two abrupt increases in error (**Figure 1**d, inset). This behavior is significant and inherent to the active learning process and underscores that the happenings in the active learning process surpass human comprehension. Third, this demonstration shows an important feature of our system: "instability avoidance," which we describe as the ability to avoid falling within a parameter domain of unstable conditions from which it becomes trapped due to the decision-making algorithm (**Figure 2**). This is analogous to repeated overcorrection in numerical analysis, which is well known by oscillations about the convergence point in the Newton Method. To further demonstrate this point, we applied our system to optimize the process conditions for a target film width and thickness using two different approaches: conventional control and real-time, active learning. In the case where conventional control was employed, which involved sequential exploration the process conditions and process-adjustment decisions based on gradient to determine how to modulate one of the current process conditions (**Figure 2**a). While conventional systems can be used for some situations, our results show that it was unable to find the optimal control conditions for the target values of width and thickness. Instead, the system became trapped as discussed earlier and failed to converge to the target value. In contrast, when using real-time active learning, the model decisively and, on occasion, drastically changed process conditions, while maintaining a balance between the unexplored regions within the process conditions and the range that was projected to be closer to the target (**Figure 2**b). In doing so, this algorithm demonstrated "instability avoidance" and efficiently converged to the target values. We interpret that this behavior was the source of the two abrupt increases in error in the root mean error plot of **Figure 1d**, inset. In fact, in this test, the conventional process control was unable to reach the target values even after 30 minutes; whereas, target values were achieved in about 8 minutes with our active learning-based autonomous control. Taken together, these autonomous optimization results demonstrate improvement in process optimization efficiency, robustness and importantly, remove the need for human intervention.

**2.2 Configuration of the Autonomous Forming System**

Next, we describe the key components of our autonomous forming system. Our autonomously-controlled forming process consists of key primary components: 1) the plastic forming component; 2) the *in-situ* evaluation component; and 3) the autonomous processor. The plastic forming component represents the mechanical material fabrication and manufacturing section of our system. This section controllably feeds plastic pellets (master batch) into a hopper, applies heat to melt and coalesce the beads to form a cohesive molten unit, extrudes this molten plastic through a slit-shaped die to form the basic cross-sectional shape, which is then cooled and collected (**Figure 1**a). Each aspect of this process: feed rate, temperature, and draw rate represent the critical interconnected parameters which must be mutually adjusted to avoid instability, such as the generation of wrinkles caused by "draw resonance phenomena", where the film exhibits fluctuations or resonances in its dimensions, as the process attempts to find a stable condition (**Figure 1**c). The *in-situ* evaluation component provides real-time diagnostics of the film dimensions (width and thickness), which then streams into the autonomous processor as a basis for the next decision (e.g. adjustment or hold) (**Figure 1**b). In this text, we denote this cycle as "characterization-adjustment" iterations. Leveraging the birefringence of plastics to vividly capture the appearance of the film, real-time detection of the formed plastic shape was made possible using a setup involving commercially available polarizers and a digital camera. Thickness information was accurately and rapidly acquired using a laser displacement meter. Together, these two evaluations afforded real-time evaluation of two key physical film dimensions, width and thickness.

The autonomous processor section represents the core of our system as it accepts process and evaluation data, interpretates this data, decides the appropriate process adjustment, and sends these orders to the plastic forming component (**Figure 1**b). As previously mentioned, due to the complex relationship between the process parameters on the resulting film, such as material grade and temperature-dependent rheological properties, pre-accumulated data for training can often be lacking. Therefore, the autonomous processor was designed using an active learning algorithm, which allows for autonomous control of the macroscopic forming process, even for materials without any pre-existing data (i.e., training data). In active learning, an untrained model is gradually trained, which progressively improves its predictive accuracy over time by interactively selecting conditions. This method of development is analogous to child development, where initially there is no knowledge and general understanding. Then, through gradual learning from various sensory inputs, the child learns numerous causal relationships, i.e. gains knowledge. In principle, provided sufficient characterization and controls, the evolved model is expected to achieve the level of a "skilled-craftsman".

In our system employing active learning, the relationship between the process conditions and the target is sequentially trained by the model. A critical step during this training is the formulation of strategies to determine the ensuing process conditions. Two key aspects here are: (1) aiming for multiple targets of different orders, such as width (~10s mm) and thickness (~sub-mm), and (2) setting the criteria to efficiently determine the next set of conditions amidst uncertain predictions with a limited number of experimental data points. In this study, we introduced the relative root mean squared error for multiple targets of different orders, transforming into a single objective function for optimization. Furthermore, we applied Bayesian optimization with an acquisition function called "Expected Improvement" to efficiently explore optimal process conditions. This metric considers both the proximity to the desired target from the data obtained thus far and the uncertainty in the regions of unexplored process conditions. By introducing these techniques into the active learning process, we achieved a closed-loop system that afforded autonomous control of the process and thereby removed the need for human intervention, while also accounting for multiple targets.

**2.3 Versatility Demonstration of Autonomous Forming Process to Diverse Target Values**

To demonstrate the ability of this system to determine process conditions for diverse target dimensions both accurately and efficiently, nine different targets were selected spanning different combinations of width and thickness. The system was directed to autonomously achieve these values similar to that described in Section 2.1 (**Figure 3**a). As a whole, target values were achieved with exceptional efficiency as evidenced by an average of 11 characterization-adjustment iterations, which corresponded to an average processing time of 19 minutes. The most rapid target required only 3 iterations (1.4 minutes), while the relatively "slowest" required 32 iterations (50 minutes). We would like to note that while 32 iterations in a 50-minute period appears long, this equates to one iteration every 1.56 minutes and exemplifies the advantages of autonomous control versus manual operation: rapid measurement-adjustment iteration and untiring ability to make data-driven decisions. To highlight the accuracy of autonomous operation, we compared the target widths and thicknesses to the achieved values (**Figure 3**b and **Figure 3**c, respectively), which show excellent agreement. The coefficient of determination ($R^2$) between the target and controlled values as a quantitative indicator of the control accuracy was estimated to be $R^2 = 0.87$ for width and 0.90 for thickness, which numerically shows the level of accuracy and importantly that our system can autonomously control the dimensions of the film with high accuracy and efficiency.

**2.4 Interpretation of the Results**

Taken together, our results demonstrate a system capable of autonomous manufacturing forming process of plastic films based on active learning to make process-adjustment decisions both accurately and efficiently to achieve targeted material dimensions. Here, we would like to discuss several aspects of our results.

As we examine the resultant process conditions for the nine target values as a whole, the complexity of the solutions from the autonomous processor can be observed (**Figure 3**d,e,f). We cannot fully explain the optimization results of a three-parameter process, because the solution is likely not unique and dependent on the initial conditions and specific trajectory during the learning process. However, in examining these results, we can make some observations from a material perspective. First, we observe that the trends of the determined process conditions for the feed rate differ from that of the draw rate and the temperature. As the target width and thickness decreases, we observe uniform and regular changes in both the draw rate and the temperature from small width and thickness to large width and thickness. This is explained by the strong sensitivity of the plastic viscosity to small (~10°C) changes in temperature. For example, while operating at the same shear rate the plastic viscosity decreases by 30% from 1315 Pa s at 270°C to 931 Pa s at 280°C (**Figure S1**). Therefore, within our range of selected target values, temperature adjustment appeared relatively minor except for the most extreme decrease in dimensionality, where it was increased to improve plastic fluidity. In contrast, the draw rate behavior shows higher speeds for decreased target widths and thicknesses, corresponding to the need for greater shear for dimensional reduction. However, we observe the most interesting behavior in the feed rate, i.e. material supply. The most glaring feature is the discontinuity across the target conditions (**Figure 3**d). We believe that this result exemplifies the unbiased decision-making process of our system, as its goal is to find a stable process condition to achieve a given target value. Despite the divergence from the other feed rate values, balance between with the viscosity (i.e., temperature) and draw rate were achieved. Thus, the system we developed can autonomously determine the unbiased process conditions appropriate for numerous target values.

**2.5 Implications and Challenges of Our Autonomous Forming Process**

The performance of our system presented here represents significant implications towards the future of the unmanned operation of continuous processes for both research and manufacturing. Specifically, our results highlight the benefits of autonomous control compared to conventional control, such as reduced measurement-adjustment iteration time, efficient condition selection, and objective decision-making (i.e., absence of human bias). Furthermore, manual operation requires the need for analysis and deliberation, which results in lost operation time; whereas autonomous control can perform these tasks nearly immediately while also affording data-driven selection of unexplored process conditions that provide the greatest increase in value. Thus, optimization can be pursued at early stages of data acquisition, even without extensive pre-existing process data.

Finally, we would like to discuss areas of improvement. While our system has demonstrated exceptional autonomous optimization, we do recognize some areas of improvement. First, higher required accuracy would require more measurement-adjustment iterations, thus longer optimization times. Therefore, for our current system, the target accuracy must be predetermined. In the future, we expect that target accuracy to be an autonomous determination by the system. Second, within our current system, the range of process conditions is predefined. This aspect is important not only from the perspective of control efficiency but also in terms of human and machine safety. For instance, in the case of this study, insufficient heating combined with high feed rate would result in excessive torque in the machine which could result in damage. In this context, we set the upper and lower limits based on the material rheological properties and the behavior of the process. We would expect that the addition of real-time system diagnostics would be a natural next step to allow the autonomous processor to independently determine parametric boundaries. In addition, autonomous control models combining domain knowledge of materials with large-scale language models is expected to represent a significant role in determining the control range for process conditions considering the nature of materials. Third, we expect that the additional process controls, such as roller temperature, ambient and humidity, and equipment geometry would add both versatility and accuracy.

## 3. Conclusion

We have demonstrated an autonomous and continuous macroscopic forming process based on an active learning decision-making algorithm. This system was demonstrated in the manufacture of plastic film forming as a model system. Equipped with real-time *in-situ* evaluation, the film properties were accurately captured and used as inputs to the autonomous processor as the measurement-adjustment iteration to realize autonomous control of process conditions governing material input, draw force, and heating, without human intervention. Moreover, it was demonstrated that even when the target width and thickness of the film were altered, the constructed system could autonomously adjust to the optimal process conditions with high precision. Importantly, due to the actively learning-based processor, zero training was required prior to assigning film dimensional targets which were achieved on an average of 11 iterations (~19 minutes). These results demonstrate the high efficiency of our system to autonomously determine process conditions without the need for human intervention. The proposed method in this study is general and can be extended to most any types of research and manufacturing processes involving continuous forming of materials from synthesis to slurries, pastes, and melts. We believe that our findings will significantly promote the autonomization of continuous processes, paving the way for broader applications and implementations in the various fields.

## 4. Experimental Section

*Material*

A commercially available polycarbonate pellet (Panlite L1225-LL, Teijin Limited) was used for the macroscopic forming process in this study. The rheological properties of this polycarbonate were measured using an oscillatory shear rheometer (AR-2000, TA Instruments, Inc.) as shown in **Figure S1**.

*Continuous Macroscopic Process*

The plastic film forming process was chosen as a typical macroscopic forming process in this study. The pellets were fed into the hopper, melted, and extruded into a T-shaped die, subsequently cooled by air and rollers to form the shape of the film. Our developed process was consisted of following five units: (1) a compounding machine (Labo-plastomill 10C100, Toyo Seiki Co., Ltd.), (2) a co-rotating twin-screw extruder unit (2D15W, Toyo Seiki Co., Ltd.) with a screw diameter of 15 mm, a screw aspect ratio of 17, and two heating blocks, (3) a T-shaped coat-hanger die (MT-60B, Toyo Seiki Co., Ltd) with a slit width of 60 mm, (4) a film haul-off unit (FT2B8, Toyo Seiki Co., Ltd.) with a thermostatic circulating oil bath (NTT-20G, Tokyo Rikakikai Co., Ltd.) to keep the roller temperature constant, and (5) a loss-in-weight feeder (AD-4826A, A&D Company, Limited.) for controlling the feed rate of the pellets to the twin-screw extruder. The appearance of our developed process is shown in **Figure S2**.

*In-situ Evaluation*

A digital single-lens reflex camera (ILCE-6600 & SEL18135, SONY Corporation) and a flat panel light-emitting diode (TH2-83X75SW, CCS, Inc.) were used to observe the appearance of the film. In addition, two linear polarizer sheets (MLPH40L-2, MeCan Imaging, Inc.) and a 1/4 wavelength sheet (MCR140N-2, MeCan Imaging, Inc.) with a retardation of 140 nm were introduced for the crossed Nicols polarization setup with a sensitive tint plate. The main optical axis of the 1/4 wavelength sheet was parallel to the flow direction of the film, while those of the linear polarizer sheets were rotated ±45 degrees from the flow direction. The edges of the film were detected by the image processing based on Hough transform (**Figure S3.**). A laser displacement meter (CL-P015, Keyence Corporation) was also placed on the roller to measure the film thickness.

*Autonomous Real-time Process Control*

To achieve autonomous process control, we introduced active learning and decision making based on acquisition function. In essence, active learning guides the autonomous optimization process by prioritizing where to try next, focusing on the most potentially informative conditions. This method is more efficient than traditional methods because it strategically selects the most valuable trials, reducing the number of trials required to optimize the conditions. A Gaussian process regressor based on the Matérn 3/2 kernel function and with a length scale of 1.0 was used to link the relationship between process conditions and target properties. The objective function was set to the negative of the relative mean squared error between the target properties (width, thickness) normalized by their target values. The reason for introducing the "relative" means squared error is to normalize the differences in the number of digits of the properties. For example, the width range is 10 – 40 mm, while the thickness range is 0 - 0.1 mm. The acquisition function for determining the next process conditions was the expected improvement (EI). This approach focuses on a balance between exploring the uncertain parameter space and exploiting the best parameter predicted by a probabilistic model from previous results. In the context of conducting real-time active learning, it is vital to ensure that the process conditions, particularly the temperature, have reached a steady state. Therefore, a situation where the process temperature falls within the range of the set value ±0.5°C is considered a balanced state. In this state, the average width and thickness of the film are calculated to serve as inputs for the active learning process. Some examples of the autonomous control in this study are shown in **Figure S4**.

*Computational Environment*

A commercial workstation was used to connect all instruments of the system and to run the autonomous control program. The specification of the workstation is Windows 10 for Workstation operating system with a central processing unit of AMD EPYC 7702P (64 cores, 128 threads, 2.0GHz clock speed), a graphics processing unit of NVIDIA GeForce RTX 3090, and 128GB of memory (eight 16GB DDR4-3200 ECC registered memories). The program of autonomous control was performed on a computational environment of python=3.9.7, cuda=11.4.152, with numpy=1.21.5, pandas=1.3.5, scikit-learn=1.0.1, matplotlib=3.5.1, opencv-python=4.5.4.60, and modAL[23].

**Acknowledgments**

We appreciate Mr. Tomohisa Hayashida, Mrs. Megumi Terauchi for their technical support of the experiments. The authors are grateful to Dr. Toshiya Okazaki, Dr. Hiroshi Morita, Dr. Ken Kokubo for their support. This work was supported by a project (JPNP16010) commissioned by the New Energy and Industrial Technology Development

Organization (NEDO).

**Conflicts of Interest**

The authors declare no conflict of interest.

**Author Contributions**

Conceptualization: S.M., K.H., Methodology: S.M., Software: S.M., T.H., Formal Analysis: S.M., Investigation: S.M., Y.M., H.N., D.N.F. Visualization: S.M., D.N.F., Supervision: K.H., Writing – Original Draft: S.M. Writing - Review & Editing: D.N.F. Specifically, S.M. designed all experimental setup, developed the software of the autonomous control, conducted experiments and calculations, and prepared this paper. T.H. developed the software for the real-time connection of multiple instruments. Y.M. contributed to the twin-screw extruder. H.N. contributed to the laser displacement meter. D.N.F. contributed to the investigation of the results, visualization, and writing the manuscript. K.H. supervised and managed the project.

**Data Availability**

Additional supporting data generated during the present study are available from the corresponding author upon reasonable request.

**Keywords:** autonomous process control, active learning, Bayesian optimization, macroscopic forming process, plastic film

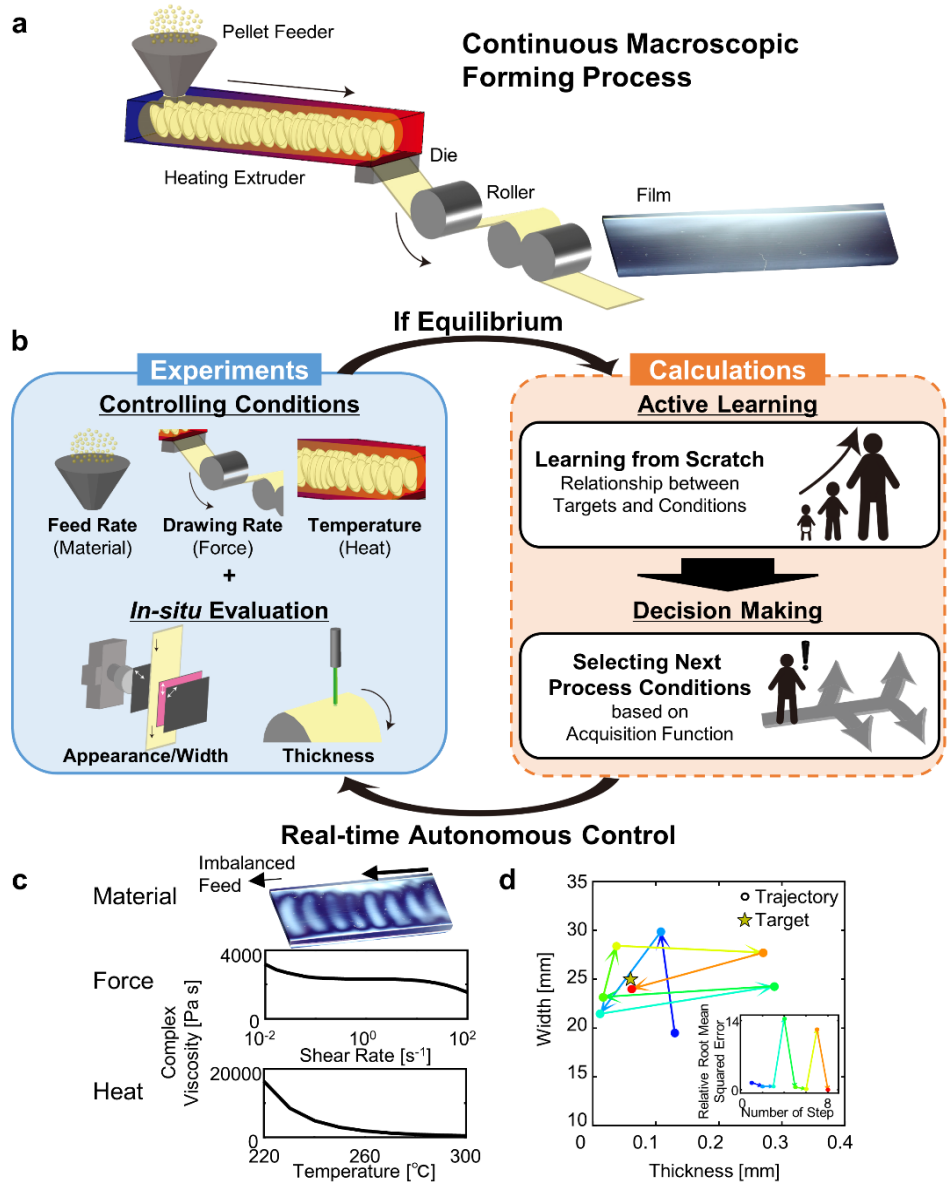

**Figure 1.** Autonomous Continuous Macroscopic Process (a) Schematic of continuous macroscopic forming process. (b) Proposed system of autonomous control of continuous macroscopic process with a closed-loop optimization of experiments with in-situ evaluation and calculations of decision making based on active learning. (c) Complexity of material, force, heat balances in continuous macroscopic forming process. (d) Trajectory of autonomous control of film dimensions (width and thickness). The inset figure represents changes in root mean squared error of targeted film dimensions.

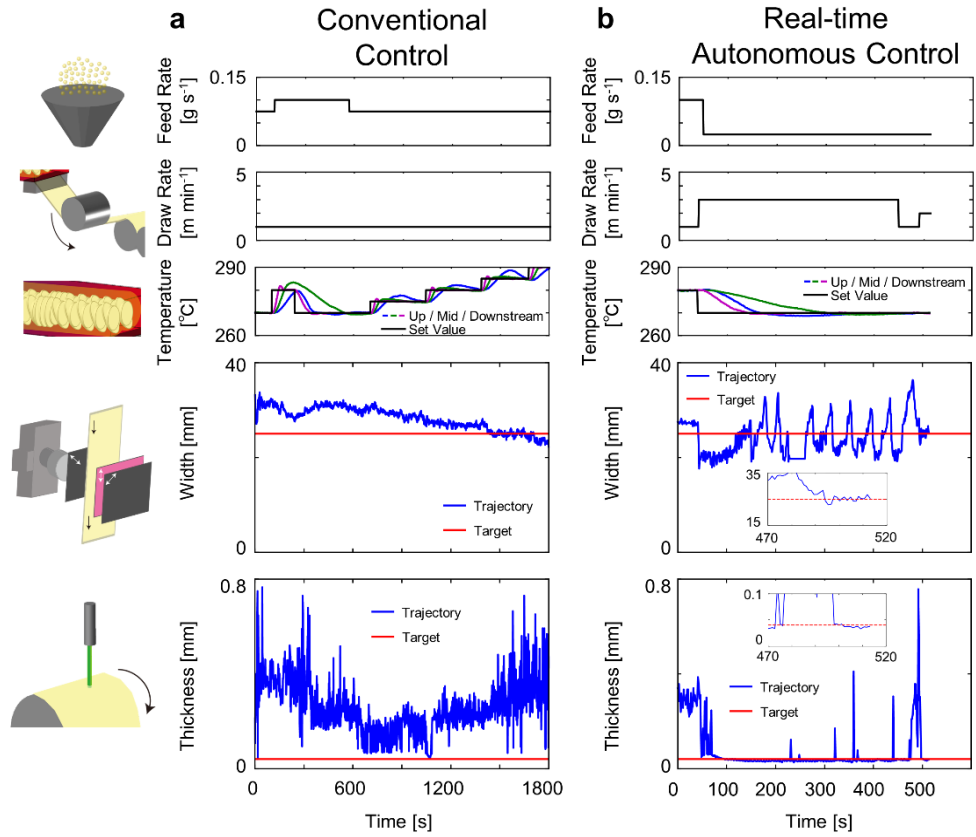

**Figure 2.** Effects of Autonomous Control. Comparison of time-series profiles of feed rate, draw rate, temperature, and film properties (width, thickness) under (a) conventional and (b) autonomous control.

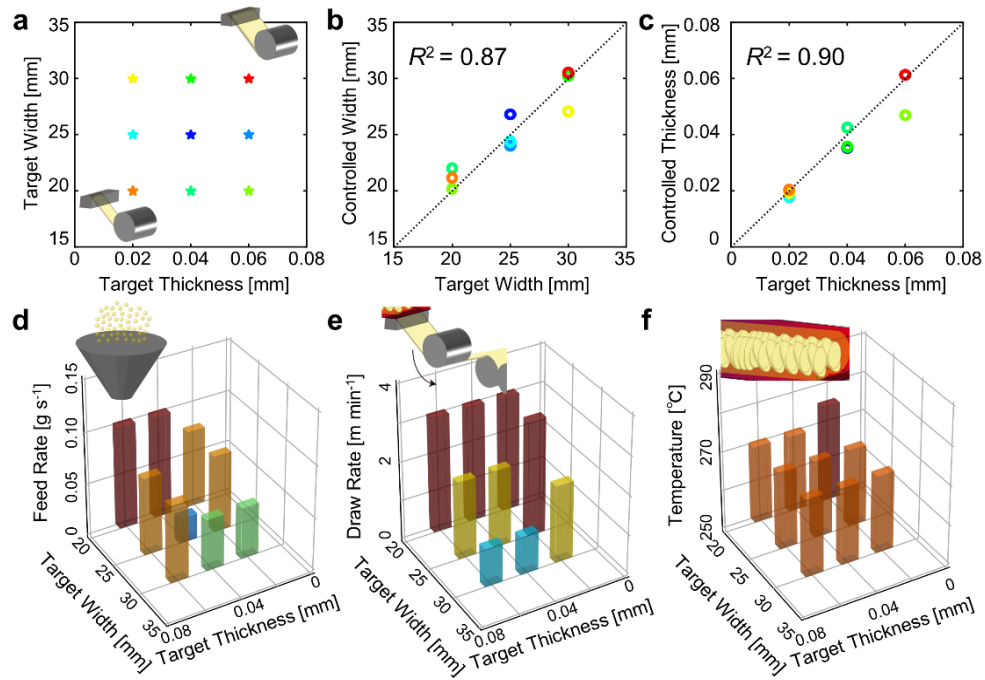

**Figure 3.** Application of Autonomous Control to Different Target Conditions. (a) A set of target width and thickness for autonomous control. Comparison of the accuracy of autonomous control of (b) width and (c) thickness. Selected conditions of (d) feed rate, (e) draw rate, () temperature for autonomous control of nine different target conditions.

Supporting Information

**Real-time Autonomous Control of a Continuous Macroscopic Process as Demonstrated by Plastic Forming**

*Shun Muroga\*, Takashi Honda, Yasuaki Miki, Hideaki Nakajima, Don N. Futaba, Kenji Hata*

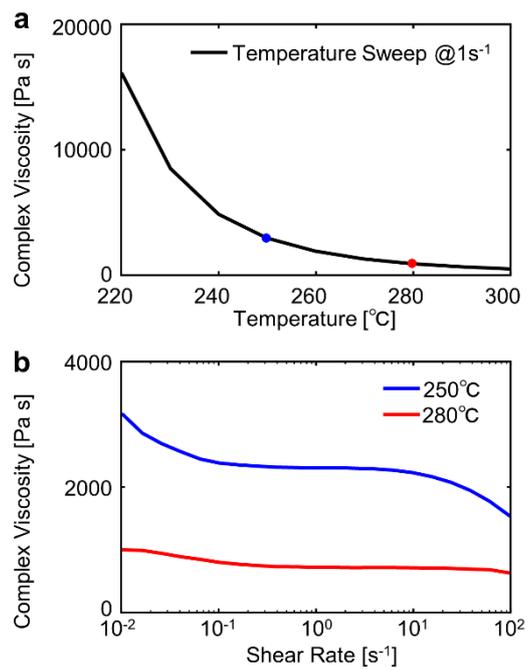

**Figure S1.** Rheological properties of the polycarbonate used in this study. (a) Temperature and (b) shear rate dependence on complex viscosity.

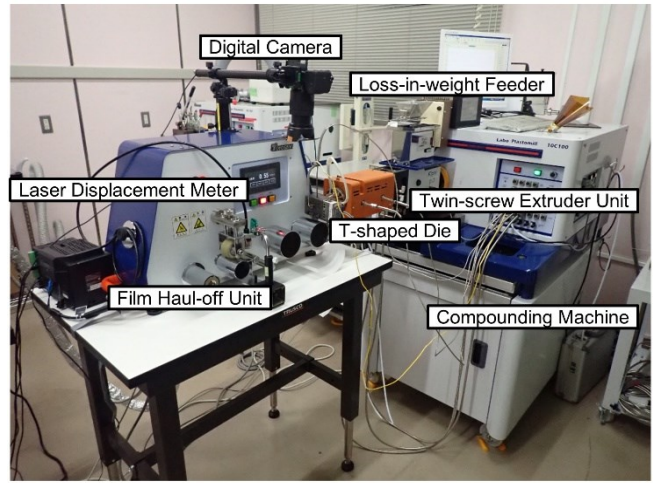

**Figure S2.** Appearance of the developed autonomous continuous macroscopic forming process.

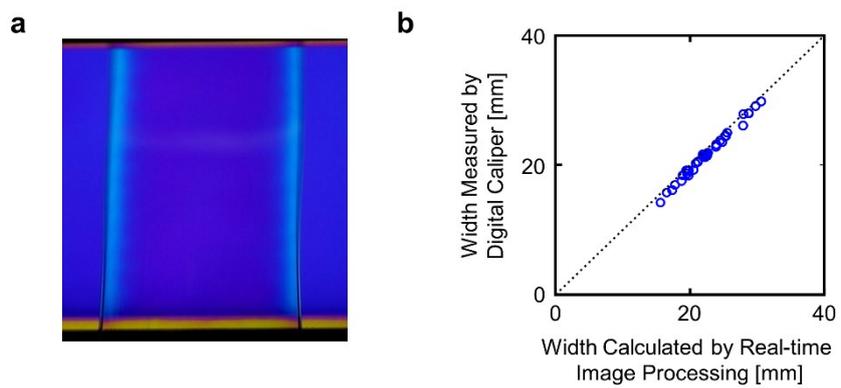

**Figure S3.** In-situ evaluation of films using real-time image processing. (a) A typical cross-Nicols polarized image of the film. (b) Comparison of the width calculated by real-time image processing and that measured by digital caliper.

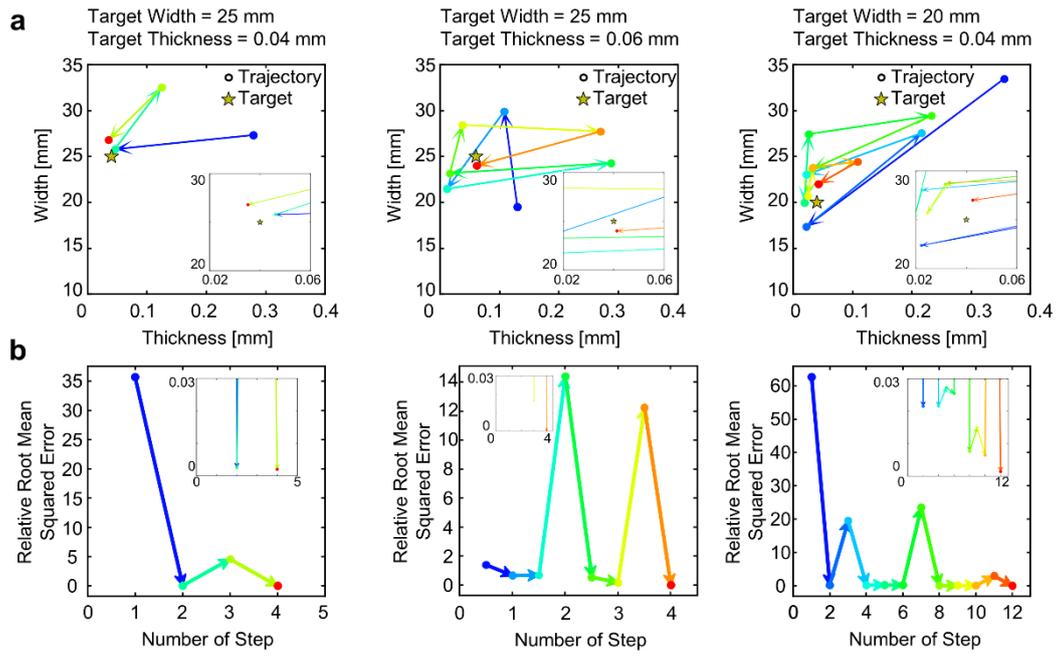

**Figure S4.** Autonomous control of different film targets. (a) Trajectory of width and thickness during autonomous control and (b) corresponding relative root mean squared error for each control step.